\documentclass[aps,pre,reprint,superscriptaddress]{revtex4-1}

\usepackage{amssymb,amsmath}

\usepackage[utf8]{inputenc}

\usepackage{epsfig}
\usepackage{bm}
\usepackage{color}
\usepackage{graphicx}
\graphicspath{{figs/}}

\def\bal#1\eal{\begin{align}#1\end{align}}

\newcommand{\Tt}{T_t}
\newcommand{\Tr}{T_r}

\newcommand{\st}{\text{st}}
\newcommand{\KK}{K}
\newcommand{\Mp}{\text{Mp}}

\newcommand{\qq}[1]{}

\begin{document}
\title{Large Mpemba-like effect in a gas of inelastic rough hard spheres}
\author{Aurora Torrente}
\affiliation{Gregorio Mill\'an Institute of Fluid Dynamics,
Nanoscience and Industrial Mathematics,
Department of Materials Science and Engineering and Chemical Engineering,
Universidad Carlos III de Madrid, 28911 Legan\'es, Spain}

\author{Miguel A. L\'opez-Casta\~no}
\affiliation{Departamento de F\'{\i}sica and Instituto de
  Computaci\'on Cient\'{\i}fica Avanzada (ICCAEx), Universidad de
  Extremadura, 06006 Badajoz, Spain}

\author{Antonio Lasanta}
\affiliation{Gregorio Mill\'an Institute of Fluid Dynamics,
Nanoscience and Industrial Mathematics,
Department of Materials Science and Engineering and Chemical Engineering,
Universidad Carlos III de Madrid, 28911 Legan\'es, Spain}

\author{Francisco Vega Reyes}
\affiliation{Departamento de F\'{\i}sica and Instituto de
  Computaci\'on Cient\'{\i}fica Avanzada (ICCAEx), Universidad de
  Extremadura, 06006 Badajoz, Spain}

\author{Antonio Prados}
\affiliation{F\'{\i}sica Te\'orica, Universidad de Sevilla, Apartado de Correos 1065, 41080
  Sevilla, Spain}
\email{prados@us.es}

\author{Andr\'es Santos}
\affiliation{Departamento de F\'{\i}sica and Instituto de
  Computaci\'on Cient\'{\i}fica Avanzada (ICCAEx), Universidad de
  Extremadura, 06006 Badajoz, Spain}


\begin{abstract}
  We report the emergence of a giant Mpemba effect in the uniformly
  heated gas of inelastic rough hard spheres: The initially hotter
  sample may cool sooner than the colder one, even when the initial
  temperatures differ by more than one order of magnitude. In order to
  understand this behavior, it suffices to consider the simplest
  Maxwellian approximation for the velocity distribution in a kinetic
  approach. The largeness of the effect stems from the fact that the rotational and
  translational temperatures, which obey two coupled evolution
  equations, are comparable. Our theoretical predictions agree very
  well with molecular dynamics and direct simulation Monte Carlo data.
\end{abstract}

\date{\today}
\maketitle



Let us consider two beakers of water at different temperatures. Mpemba
and Osborne showed that the initially hotter sample cools sooner under
certain conditions~\cite{mpemba_cool_1969},
i.e., the curve giving the time evolution of its temperature crosses
that of the initially cooler sample and stays below it for longer times.
This is called the Mpemba memory effect, which is known since
antiquity in cultures for which water in the form of ice and snow is
common~\cite{aristotle_works_1931}. Later, the Mpemba effect has
been clearly identified in different physical systems
\cite{greaney_mpemba-like_2011,Ahn16,lu_nonequilibrium_2017,Lasanta17,
  lasanta_spinGlasses_2018}, although there is still some debate about
its existence in water \cite{Burridge16}.

From a physical point of view, one would like to answer
how different the initial preparation of two samples of the system
under study must be so that the Mpemba effect arises. This is the main---currently unresolved in
general---question,  although there has been some recent progress in this
respect~\cite{lu_nonequilibrium_2017,Lasanta17}. Lu and
Raz~\cite{lu_nonequilibrium_2017} analyzed the Mpemba effect in a
generic Markovian system by monitoring the relaxation of an
entropy-like variable that measures the distance to the steady
state. This makes it possible to define and investigate Mpemba-like
effects in systems for which there is not an obvious definition of a
nonequilibrium temperature, but makes  the comparison
with the usual experimental setup described above difficult.

A different approach was carried out by some of us in the study of the
Mpemba effect for a granular fluid of \textit{smooth} hard
spheres~\cite{Lasanta17}.  Therein, the granular
temperature---basically the average kinetic energy per particle---is the physical
quantity monitored to investigate the Mpemba effect. In the smooth-sphere
case, the angular velocities play no role since there is no energy
transfer between the translational and the rotational degrees of
freedom, and the kinetic energy is thus purely translational.  We
showed that the Mpemba effect stems from the coupling of the granular
temperature and the kurtosis, which measures the deviation of the
velocity distribution function from the Maxwellian shape at the lowest
order. More specifically, it is the difference between the initial
values of the kurtosis of the two samples that controls the appearance
of the Mpemba effect.

In the granular fluid of smooth hard spheres, the kurtosis is
typically small. On the one hand, this facilitates the theoretical
analysis, because it makes it possible to linearize the evolution
equations and thus give a quantitative prediction of how different the
initial kurtoses must be to enable the Mpemba effect. On the other
hand, the smallness of the kurtosis limits the magnitude of the Mpemba
effect: The initial temperatures must be quite close; see Eq.~(5) and Fig.~1(b) of Ref.~\cite{Lasanta17}.

It has very recently been shown that a different memory effect, the
Kovacs effect
\cite{kovacs_isobaric_1979,bertin_kovacs_2003,mossa_crossover_2004,aquino_kovacs_2006,prados_kovacs_2010},
is much larger and more complex in a granular gas of \emph{rough}
spheres~\cite{lasanta_emergence_2018} than in the smooth-sphere case~\cite{prados_kovacs-like_2014,trizac_memory_2014}.  What makes it
possible to understand the largeness of the Kovacs effect is the
coupling between the translational and rotational temperatures, which
are of the same order of magnitude. In addition, the basic features of
the memory effect can be understood within the Maxwellian (Gaussian)
approximation, without having to resort to higher order cumulants.

The above picture prompts us to look into the Mpemba effect in a fluid
of rough inelastic hard spheres. Remarkably, we show in the following
that the Mpemba effect can be explained within a Gaussian framework,
both qualitatively and quantitatively. Physically speaking, the
coupling between the rotational and translational degrees of freedom
and thus the existence of two comparable but different temperatures
suffices to explain the memory effect.  Moreover, we give a picture of
the underlying physical conditions and discuss the possible relevance
of the two-temperature mechanism in other systems.





Therefore, let us consider a dilute gas of inelastic rough hard
spheres, with mass $m$, diameter $\sigma$, and moment of inertia $I$.
Henceforth, we employ the dimensionless moment of inertia
$\kappa\equiv 4I/m\sigma^{2}$, and its specific value for uniform
solid spheres ($\kappa=\frac{2}{5}$) whenever a definite value is
needed. Translational and angular particle velocities are denoted as
$\bm{v}$ and $\bm{\omega}$, respectively.

Collisions between macroscopic particles are inelastic, i.e., energy
is not conserved~\cite{FLCA93}. For the inelastic rough hard sphere
model, a binary collision is characterized by two parameters: the
coefficient of normal restitution $0\leq\alpha\leq 1$ and
the coefficient of tangential restitution
$-1\leq\beta\leq 1$~\cite{K10a, brilliantov_kinetic_2004,G19}. They are
intrinsic properties of the material and determine the shrinking of
the normal and tangential components of the relative velocity of the
two surface points at contact~\cite{note4}.  The collisional model based on these two
parameters is sufficiently accurate in a variety of
materials~\cite{FLCA93}, with most of them presenting experimental
values in the intervals $\alpha\in(0.7,0.95)$ and $\beta\in(-0.5, 0.5)$
\cite{L99}. Consistently, the analysis carried out in this paper focuses on this region of the
$(\alpha,\beta)$ plane.

Additionally, energy is homogeneously injected to the translational
degrees of freedom of all the grains by a \emph{stochastic thermostat}
$\mathbf{F}$ modeled as a Gaussian white noise, i.e.,
$\langle {\bf F}_i(t) \rangle ={\bf 0}$,
$\langle {\bf F}_i(t) {\bf F}_j(t') \rangle =\mathsf{I}m^2 \chi_0^2
\delta_{ij}\delta(t-t')$, where $i,j$ refer to the particles' indexes,
$\mathsf{I}$ is the $3\times 3$ unit matrix, and $\chi_0^2$ gives the
``strength'' of the stochastic forcing~\cite{W96,WM96,SBCM98,vNE98,MS00,GSVP11a}.


Here, we provide the minimal theoretical framework needed for the
understanding of the Mpemba effect in the granular gas (see
Refs.~\cite{reyes_role_2014,Vega15,supplemental} for a detailed
account of the kinetic theory calculations). The dynamics of our
system is governed by the inelastic Boltzmann--Fokker--Planck equation
for the single-particle velocity distribution function
$f(\bm{v},\bm{\omega},t)$ \cite{MS00}. From the kinetic equation, the
evolution equations for the average quantities of interest are
derived~\cite{reyes_role_2014,Vega15}.

We restrict ourselves to homogeneous and isotropic states, for which
$\langle\bm{v}\rangle=\mathbf{0}$ and
$\langle\bm{\omega}\rangle=\mathbf{0}$.  The basic physical
information is thus encoded in the translational and rotational granular
temperatures $\Tt={\frac{m}{3}}\langle v^2\rangle$ and
$\Tr={\frac{I}{3}}\langle \omega^2\rangle$. Alternatively, the same
information is provided by the temperature ratio $\theta$ and the
total temperature $T$,
\begin{equation}
\label{theta,T}
\theta(t)=\frac{\Tr(t)}{\Tt(t)}, \quad T(t)=\frac{\Tt(t)+\Tr(t)}{2}.
\end{equation}
The granular gas is inherently a nonequilibrium system and, therefore, equipartition is broken, i.e.,
 $\theta\neq 1$. Thus, the
simplest description of the rough-sphere granular gas is provided by the
Maxwellian approximation, in which the following bivariate
Gaussian form is assumed for the velocity distribution function,
$f(\bm{v},\bm{\omega},t)\simeq n\!\left[{mI}/{4\pi^{2}\Tt(t)\Tr(t)}
  \right]^{\frac{3}{2}}
  \exp\left[-\frac{mv^{2}}{2\Tt(t)}-\frac{I\omega^{2}}{2\Tr(t)}\right]$,
where $n$ is the number density.

In the long-time limit, the granular gas reaches a steady state due to
the action of the stochastic force. This steady state is completely
characterized by $\theta^{\st}$ and $T^{\st}$ in the Maxwellian
approximation%
. Their
expressions in terms
of the coefficients of restitution and the stochastic forcing
intensity are \cite{Vega15}
\begin{subequations}\label{eq:theta-gamma-st}
\begin{align}
\theta^{\st}=&\frac{1+\beta}{2+\kappa^{-1}(1-\beta)}, \quad T^{\st}=\frac{1+\theta^\st}{2}\left(\frac{3m^{3/2}\chi_{0}^{2}}{4\sqrt{\pi}n\sigma^{2}\gamma^{\st}}\right)^{2/3}, \\
\gamma^{\st}\equiv& 1-\alpha^{2}+\frac{2(1-\beta^{2})}{2+\kappa^{-1}(1-\beta)}.
\end{align}
\end{subequations}
Note that
$\theta^{\st}\leq 1$ is independent of $\alpha$ in the Maxwellian
approximation but higher order approximations introduce a---rather
weak---dependence on $\alpha$ \cite{Vega15}.

It is useful to introduce dimensionless variables for temperature and
time. Then, we define $T^{*}\equiv T/T^{\st}$ and
$t^{*}\equiv 2n\sigma^{2}\sqrt{\pi T_{t}^{\st}/m}t$. In the Maxwellian
approximation, $T^{*}$ and $\theta$ evolve according
to~\cite{supplemental}
\begin{equation}
\label{A1}
\partial_{t^{*}}\ln T^*=\Phi(T^*,\theta), \quad
\partial_{t^{*}}\ln\theta=\Psi(T^*,\theta),
\end{equation}
with the definitions
 \begin{subequations}\label{Phi-Psi}
\begin{align}
\label{Phi}
\Phi(T^*,\theta)=&\Phi_1(T^*)+\Phi_2(T^*,\theta)+\Phi_3(T^*,\theta),
\\
\label{Psi}
\Psi(T^*,\theta)=&-(1+\theta)\left[\Phi(T^*,\theta)-
              \frac{\Phi_3(T^*,\theta)}{\theta(1-\theta^\st)}\right].
\end{align}
\end{subequations}
Above, we have introduced the notation
\begin{subequations}
\label{Phi-1-2-3}
\begin{align}
\Phi_1&\equiv \frac{2}{3}\frac{\gamma^\st}{T^*(1+\theta^\st)}, \quad
\Phi_2\equiv-
\frac{2}{3}\sqrt{\frac{T^*(1+\theta^\st)}{\left(1+\theta\right)^3}}\gamma^\st, \\
\Phi_3&\equiv
\frac{2}{3}\KK\sqrt{\frac{T^*(1+\theta^\st)}{\left(1+\theta\right)^3}}\left(1-\frac{\theta}{\theta^\st}\right)(1-\theta^\st),
\end{align}
\end{subequations}
with $\KK\equiv \kappa (1+\beta)^{2}/(1+\kappa)^{2}$~\cite{note7}.
Note that the time evolution of the temperature is governed by the
function $\Phi$, which  does not only depend on $T^{*}$; this is a
necessary condition for the appearance of the Mpemba effect. 

Imagine
two initial states $(T_{0A}^*,\theta_{0A})$ and
$(T_{0B}^*,\theta_{0B})$, with $T_{0A}^*>T_{0B}^*>1$, of the same
granular gas, i.e., with the same values of the coefficients of
restitution $\alpha$ and $\beta$. Let us denote by $T_A^*(t^*)$ and
$T_B^*(t^*)$ the associated decays of the temperature to the
steady state: A Mpemba-like effect is brought about when there exists
a crossing time $t^*_\times$ such that $T_A^*(t^*)<T_B^*(t^*)$ for
$t^*>t^*_\times$.

A necessary---and physically intuitive---condition for having the Mpemba
effect is that the initially hotter sample cools faster than the
cooler one for short times, when the system still keeps ``memory'' of
its initial conditions and is in the first stage of the so-called
kinetic regime. For short enough times, we can consider that the
system is exponentially cooling with a characteristic rate roughly
equal to the initial value of $-\Phi$ \cite{note12} and then a necessary condition for the  Mpemba effect to be
present is
\begin{equation}\label{eq:phiA-phiB}
  \Phi(T_{0A}^*,\theta_{0A})<  \Phi(T_{0B}^*,\theta_{0B}).
\end{equation}

Let us investigate the behavior of $\Phi(T^{*},\theta)$ as a function of
$\theta$, for fixed $T^{*}$, to understand under which conditions the
Mpemba effect is expected. There are three distinct terms in $\Phi$:
(i) the first one, $\Phi_{1}(T^{*})$, is a \textit{heating term} that
stems from the stochastic forcing and is thus independent of $\theta$,
(ii) the second one, $\Phi_{2}(T^{*},\theta)$, is the typical
\textit{cooling term} of granular gases, which is also present for
smooth spheres~\cite{note6}, and (iii)
the third one, $\Phi_{3}(T^{*},\theta)$, is a purely \textit{roughness
  term} (note that $\KK=0$ for $\beta=-1$) and heats (cools) the system
when $\theta<\theta^{\st}$ ($\theta>\theta^{\st}$). The sign and magnitude of
$\Phi(T^*,\theta)$ results from the competition among those three
terms.

In light of the above, we analyze the behavior of
$\Phi_{2}$ and $\Phi_{3}$ as a function of $\theta$, for fixed
$T^{*}$. While the cooling term $\Phi_{2}$ is a monotonically increasing
function of $\theta$~\cite{note8}, the roughness
term $\Phi_{3}$ shows a more complex behavior. Starting from
$\theta= 0^{+}$, $\Phi_{3}$ first decreases with increasing $\theta$,
vanishes at $\theta=\theta^\st$, reaches a (negative) minimum value at
$\theta=2+3\theta^\st$, and finally tends to zero from below in the
limit $\theta\to \infty$.

\begin{figure}
    \includegraphics[height=1.725in]{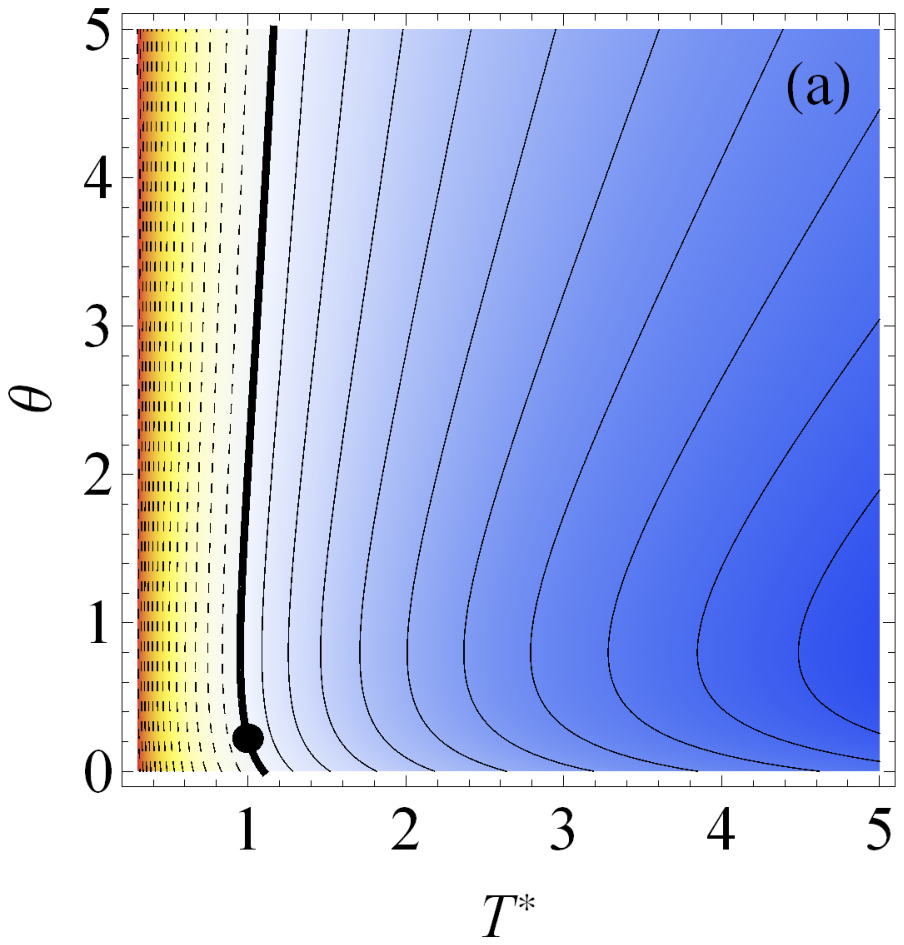}
    \includegraphics[height=1.71in]{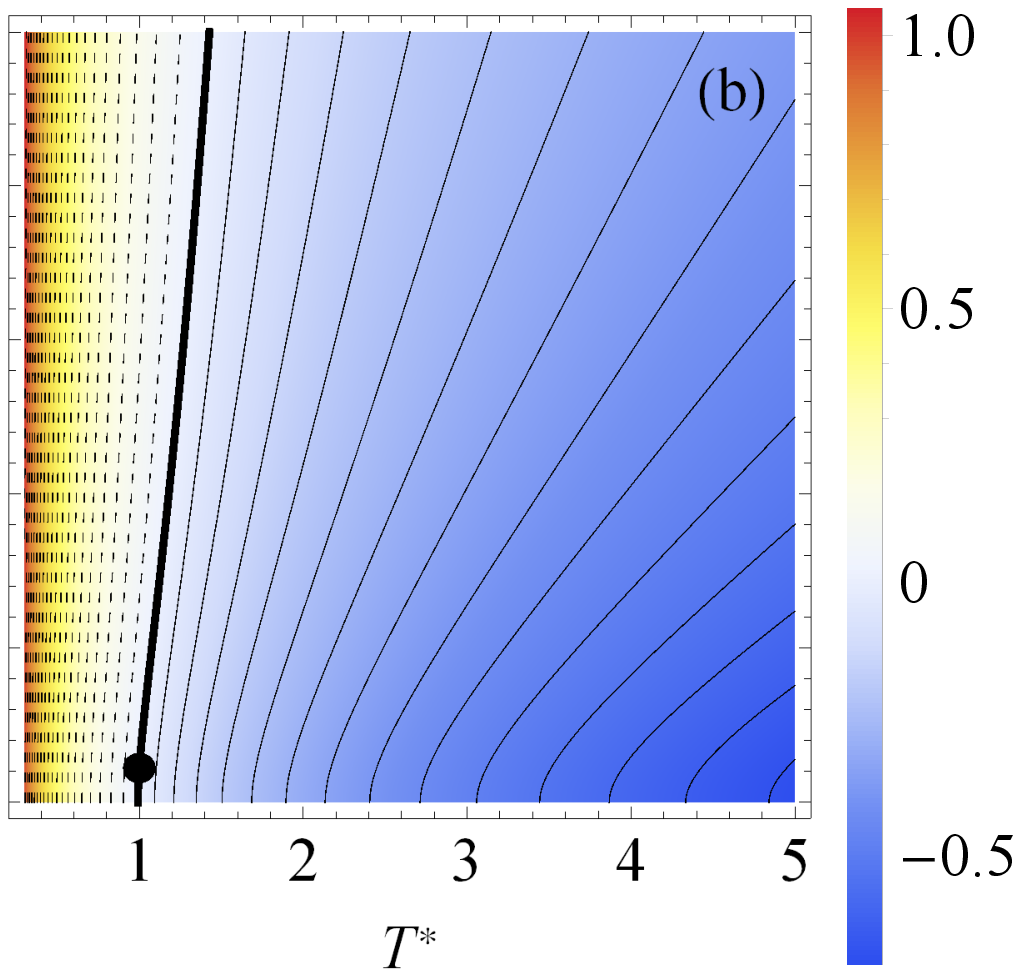}
  \caption{Density plots of $\Phi(T^*,\theta)$ as defined by
    Eq.~\eqref{Phi}. Two representative examples of the coefficient of
    restitution, (a) $\alpha=1$ and (b) $\alpha=0.9$, are considered,
    for $\beta=0$. The contour lines (solid for negative $\Phi$,
    dashed for positive $\Phi$) are separated by an amount
    $\Delta\Phi=0.05$. The thick solid line is the locus
    $\Phi(T^*,\theta)=0$ and the circle marks the steady-state point
    ($T^*,\theta)=(1,\theta^\st)$.  }
\label{Tdot}
\end{figure}

The overall behavior of $\Phi_2+\Phi_3$ as a function of $\theta$
depends on the values of the coefficients of restitution
$(\alpha,\beta)$.  Both $\Phi_2$ and $\Phi_3$ grow with increasing
$\theta$ beyond the minimum of $\Phi_3$, i.e., for
$\theta>2+3\theta^{\st}$. On the other hand, as $\alpha$ approaches
unity, the decay of $\Phi_{3}$ for small $\theta$ dominates over the
growth of $\Phi_{2}$, resulting in a nonmonotonic dependence of $\Phi$
on $\theta$. This is illustrated in Fig.~\ref{Tdot}(a), which puts
forward a density plot of $\Phi$ as a function of $(T^{*},\theta)$ for
uniform solid spheres in the limiting case $(\alpha=1,\beta=0)$.  A
nonmonotonic behavior is neatly observed, especially to the right of
the locus $\Phi(T^{*},\theta)=0$, i.e., where $\Phi<0$ and the system
cools. As $\alpha$ is decreased, the magnitude of the cooling term
$\Phi_{2}$ increases, eventually becoming the dominant one for small
$\theta$ if $1-\alpha$ is large enough. Therein, a monotonically
increasing behavior is observed, as illustrated in Fig.~\ref{Tdot}(b)
for $(\alpha=0.9,\beta=0)$.

\begin{figure}[tbp]
  \includegraphics[width=2.25in]{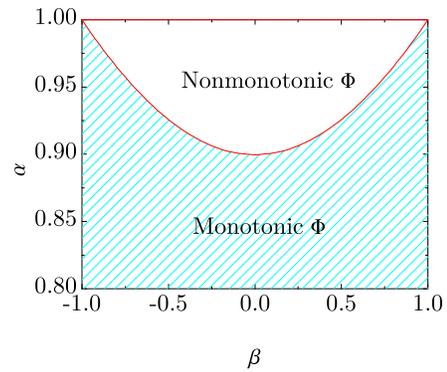}
\caption{Locus $\widetilde{\theta}(\alpha,\beta)=0$ in the
  $(\beta,\alpha)$ plane. We have a nonmonotonic behavior of
  $\Phi(T^*,\theta)$ vs $\theta$ with a minimum at
  $\widetilde{\theta}>0$ above the curve, where $\widetilde{\theta}$ is
  given by Eq.~\eqref{thetamin}, whereas $\Phi(T^*,\theta)$ has a
  monotonically increasing behavior below the curve. Here,
  $\kappa=\frac{2}{5}$ (uniform solid spheres).}.
\label{alpha_vs_beta}
\end{figure}

To carry out a more quantitative analysis, we impose that $\partial\Phi(T^*,\theta)/\partial
\theta|_{\widetilde{\theta}}=0$, which leads to
\begin{align}
  \label{thetamin}
\widetilde{\theta}(\alpha,\beta)=
2-3\kappa-3(1+\kappa)\frac{1-\alpha^2}{1-\beta^2}
\end{align}
and study the sign of $\widetilde{\theta}$ \cite{note13}. If
$\widetilde{\theta}\leq 0$, there is no physically meaningful minimum
and $\Phi$ is a monotonically increasing function of $\theta$. If $\widetilde{\theta}>0$, $\Phi$ displays a minimum at $\theta=\widetilde{\theta}$.
Equation~\eqref{thetamin} implies that $\widetilde{\theta}>0$ if
$\alpha$ is sufficiently close to unity and $|\beta|$ is sufficiently
small. This is consistent with the qualitative discussion above, and
it is illustrated in Fig.~\ref{alpha_vs_beta}. Therein, the locus
$\widetilde{\theta}=0$
separates the
regions inside which $\Phi$ is monotonic (below it) and nonmonotonic
(above it). On the one hand, $\widetilde{\theta}\leq 0$ for all
$\beta$ when $\alpha\leq\alpha_{c}=\sqrt{(1+6\kappa)/3(1+\kappa)}$,
which gives $\alpha_{c}=\sqrt{17/21}\simeq 0.9$ for uniform solid
spheres. On the other hand, $\widetilde{\theta}>0$ for
$\alpha>\alpha_c$ only if
$\beta^2<1-3(1-\alpha^2)(1+\kappa)/(2-3\kappa)$~\cite{note9}.



\begin{figure}
    \includegraphics[width=\columnwidth]{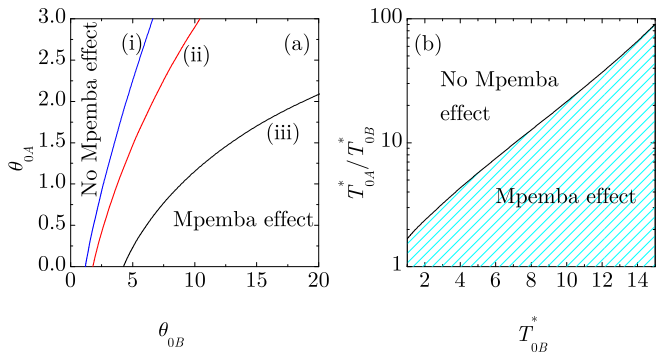}
  \caption{Phase diagrams of the Mpemba effect for $\alpha=0.9$ and
    $\beta=0$. In panel (a), we consider the
    $(\theta_{0B},\theta_{0A})$ plane: The Mpemba effect is present
    (absent) for points below (above) the plotted lines, which
    correspond to several choices of the initial conditions;
    specifically, (i) $(T_{0A}^*,T_{0B}^*)=(5,4)$, (ii)
    $(T_{0A}^*,T_{0B}^*)=(4,3)$, and (iii)
    $(T_{0A}^*,T_{0B}^*)=(5,3)$. Note the different scales in both
    axes. In panel (b), we plot the line in the
    $(T_{0B}^{*},T_{0A}^{*}/T_{0B}^{*})$ plane below which the Mpemba
    effect may appear, provided that the kinetic energy of the
    initially hotter (colder) sample is concentrated in the
    translational (rotational) degrees of freedom to a sufficient
    extent.  }
\label{phase_diag}
\end{figure}

For the sake of conciseness, we restrict ourselves to the simpler
monotonic situation $\alpha\leq\alpha_{c}$ in the remainder of the
paper. Let us look again at Fig.~\ref{Tdot}(b), in
which the limiting---less favorable---case
$(\alpha=\alpha_{c},\beta=0)$ is shown. As already stated before, we consider two
points,  $A\equiv(T_{0A}^{*},\theta_{0A})$ and
$B\equiv(T_{0B}^{*},\theta_{0B})$  with
$T_{0A}^{*}>T_{0B}^{*}$,
corresponding to different initial conditions, to study the Mpemba effect. Since $\Phi$ is monotonic, it suffices to
take $\theta_{0A}<\theta_{0B}$ (i.e., the initially hotter system has
its kinetic energy more concentrated in the translational modes than
the initially cooler one) to fulfill
Eq.~\eqref{eq:phiA-phiB}.
Equation~\eqref{eq:phiA-phiB}, however, is not a sufficient condition
for the Mpemba effect to appear: The disparity between $\Phi(T_{0A}^*,\theta_{0A})$ and
$\Phi(T_{0B}^*,\theta_{0B})$ must be large enough, because the relaxation is not purely
exponential. This entails that the disparity between $\theta_{0A}$ and
$\theta_{0B}$ must be also large enough. This is illustrated in Fig.~\ref{phase_diag}(a), where we analyze the emergence of the
Mpemba effect for several pairs of the initial temperatures
$(T_{0A}^{*},T_{0B}^{*})$. For each one of these pairs, the lines in the
$(\theta_{0B},\theta_{0A})$ plane delimiting the regions with  and
without  the Mpemba effect are plotted.

For the emergence of the Mpemba effect, the most favorable situation
is having the kinetic energy completely concentrated in (i) the
translational degrees of freedom for the higher temperature
$T_{0A}^{*}$, i.e., $\theta_{0A}=0$, and (ii) the rotational degrees of
freedom for the lower temperature $T_{0B}^{*}$, i.e.,
$\theta_{0B}\to\infty$. This limiting case is considered in Fig.~\ref{phase_diag}(b). The range of initial temperature ratios
$T_{0A}^{*}/T_{0B}^{*}$ leading to the emergence of the Mpemba effect
increases (almost exponentially) with the colder initial temperature
$T_{0B}^{*}$. For instance, if $T_{0B}^{*}=15.3$, $T_{0A}^{*}$ can be $100$ times
larger than $T_{0B}^{*}$ and still the Mpemba effect is observed. In
all the cases plotted in Fig.~\ref{phase_diag}, we have considered
that the Mpemba effect is present when the crossing time $t^*_\times$---if it ever exists---is
smaller than $10$~\cite{note10}.


To check the accuracy of our theoretical predictions, we have carried
out molecular dynamics (MD) and direct Monte Carlo (DSMC)
simulations~\cite{supplemental}.  Different systems are considered by
varying the values of $\alpha$ and $\beta$. For each particular pair
$(\alpha,\beta)$, the system is initialized at three points $A$, $B$,
and $C$ with $(T^*_{0A},\theta_{0A})=(5,0.01)$,
$(T^*_{0B},\theta_{0B})=(4,10)$, and $(T^*_{0C},\theta_{0C})=(3,100)$,
respectively. According to our theoretical framework, the Mpemba
effect is expected to emerge in the cases $(A,B)$ and $(A,C)$, but not
in the case $(B,C)$.

\begin{figure*}
           \includegraphics[width=1.5\columnwidth]{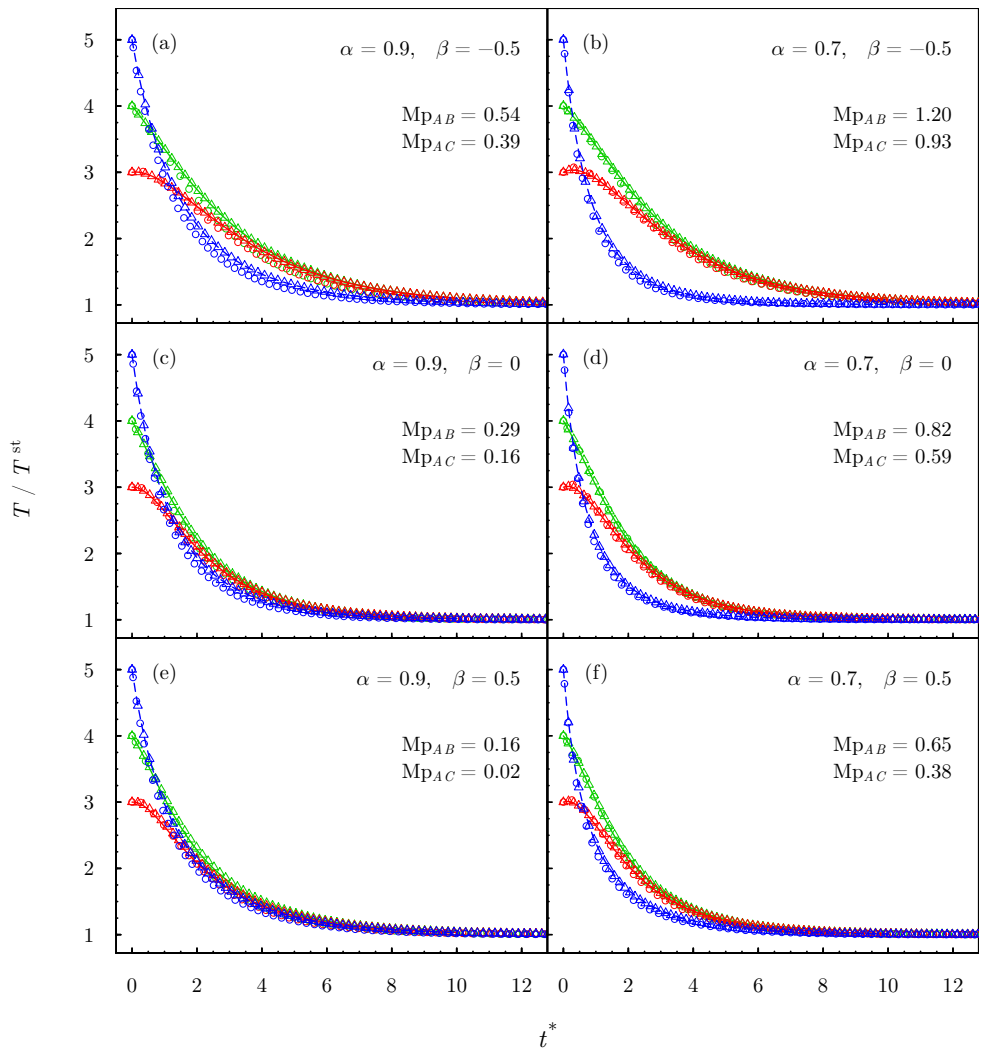}
           \caption{Mpemba effect in the granular gas of rough hard
             spheres. The relaxation of the scaled temperature $T^{*}$
             to its steady-state value is shown for different values
             of the coefficients of restitution $\alpha$ and
             $\beta$. The MD data (open circles) and, especially, the
             DSMC data (triangles) are in very good agreement with the
             theoretical values (lines). Dashed blue, solid green
             (light gray), and solid red lines refer to initial states
             given by $(T_{0A}^*,\theta_{0A})=(5,0.01)$,
             $(T_{0B}^*,\theta_{0B})=(4,10)$, and
             $(T_{0C}^*,\theta_{0C})=(3,100)$, respectively. The
             Mpemba effect is neatly observed, with the initially
             hottest system being the first one to reach the steady
             state.  }
\label{MD}
\end{figure*}

Figure~\ref{MD} shows the time evolution of the granular
temperature. The curves correspond to $\alpha=0.7$ and $0.9$, and
$\beta=-0.5$, $0$, and $0.5$, but the qualitative picture explained
below remains unaltered for other choices, as long as
$\alpha\leq\alpha_{c}\simeq 0.9$. The Mpemba effect is clearly
observed as the initially hottest sample ($A$) is the fastest one to
reach the steady state. Moreover, in all the cases, the three plotted
curves---the MD data (circles), the DSMC data (triangles), and the
Maxwellian theory (line)---present a remarkably excellent agreement.

Qualitatively, the strength of the Mpemba effect---the separation
between the two relaxation curves for times longer than
$t_{\times}^*$---increases as $\alpha$ decreases at fixed $\beta$ (higher
inelasticity) and as $\beta$ goes to smaller values at fixed $\alpha$
(lower roughness). Thus, quantitatively, we can measure the magnitude
of the Mpemba effect by defining a \textit{Mpemba parameter}
$\Mp_{\!AB}$ as the extremum of the difference of dimensionless
temperatures $T_{B}^{*}(t)-T_{A}^{*}(t)$ for $t^*>t_{\times}^*$. The values of $\Mp_{\!AB}$ and $\Mp_{\!AC}$ obtained from the solution of Eq.\ \eqref{A1} are shown in each one of the panels of Fig.\ \ref{MD} and are consistent with the qualitative
behavior described above. For fixed roughness $\beta$, the more
inelastic the system is, the larger $\Mp$ becomes. For fixed
inelasticity $\alpha$, the smoother the system is, the larger $\Mp$
becomes. Note that the values of $\Mp$ are typically of the order of unity for
our system of rough spheres. In the smooth-sphere case, however, $\Mp$ is much
smaller, being typically of the order of a few thousandths for the cases
reported in Ref.~\cite{Lasanta17}. While for smooth spheres the influence of the kurtosis on the evolution of $T^*$ is very weak, a strong impact of the rotational-translational partition is present in the case of rough spheres.



We have neatly observed the Mpemba effect in the
gas of inelastic rough hard spheres. It stems from the coupling in the
evolution of the rotational and translational temperatures, which are,
in general, of the same order, $\theta=O(1)$. For low enough $\alpha$,
i.e., large enough inelasticity, we have shown that the more
concentrated in the translational degrees of freedom the total kinetic
energy is, the faster the system cools.
Therefore, the initially hotter system must have its kinetic energy
more concentrated in the translational modes than the initially cooler
one to facilitate the Mpemba effect. Moreover, we have quantified how
large this concentration must be. In the rough-sphere case, the initial
temperatures of the two samples can be quite different and definitely
do not need to be close to each other, as happens in the smooth-sphere granular gas.

The strength of the effect has been quantified by the Mpemba parameter
$\Mp$ defined as the maximum separation---relative to the stationary
temperature---between the relaxation curves after the crossing. The
Mpemba effect in the rough-sphere granular gas is really huge, with
typical order of unity values of $\Mp$. This has to be contrasted with
the Mpemba effect in the smooth-sphere case that, although distinctly
observed in Ref.~\cite{Lasanta17}, is quite small ($\Mp\sim 10^{-3}$).

Also, it is worth emphasizing that the Mpemba effect can be explained
with a minimal Gaussian approximation for the distribution
function. The largeness of the effect shows that the system is
sweeping far-from-equilibrium states but, interestingly, these states
can be described by a sort of ``local equilibrium'' approximation but
with two distinct temperatures. This is at difference with the
hydrodynamic regime, in which the rotational temperature becomes
enslaved to the total one for long enough times, i.e., $\theta$
becomes time independent over the hydrodynamic scale and $T$ is the
relevant hydrodynamic field.

In the rough-sphere case, the corrections introduced by the cumulants are
expected to remain small~\cite{note11}.  This expectation is supported
by the exceptionally good agreement found in Fig.~\ref{MD} between the
Maxwellian analytical predictions and the DSMC data---which give the
numerical integration of the kinetic equation. In addition, the
agreement of our theory with MD simulations tells us that our kinetic
description holds for the low-density gas.

The results in the present work suggest that the two-temperature
mechanism found here may be significant for observation of a huge---and
thus experimentally measurable---Mpemba effect in a variety of systems
with several distinct temperatures, such as structural
glasses~\cite{narayanaswamy_model_1971,debolt_analysis_1976}, metals
undergoing plastic
deformation~\cite{roy_chowdhury_non-equilibrium_2019}, and granular
mixtures~\cite{dahl_kinetic_2002}.  This is enough to bring to the
fore the Mpemba effect, which may even be enhanced if other variables
like higher order cumulants are also relevant.

This work has been supported by the Spanish
Ministerio de Ciencia, Innovaci\'on y Universidades and the Agencia
Estatal de Investigaci\'on Grants (partially financed by the ERDF)
No.~MTM2017-84446-C2-2-R (A.L. and A.T.), No.~FIS2016-76359-P
(F.V.R. and A.S.), No.~FIS2017-84440-C2-2-P (A.T.), and
No.~PGC2018-093998-B-I00 (A.P.), and by the Junta de Extremadura
(Spain) Grants No.~IB16087 and No.~GR18079 (F.V.R., A.S., and M.L.C.),
partially funded by the ERDF.  Use of computing facilities from
Extremadura Research Center for Advanced Technologies (CETA-CIEMAT),
funded by the ERDF, is also acknowledged.

\bibliographystyle{apsrev}


\bibliography{MRWN_R2}

\begin{thebibliography}{44}
\bibitem[{not({\natexlab{a}})}]{note4}
\bibinfo{note}{More specifically, the normal and tangential components of the
  postcollisional velocity are equal to their precollisional counterparts
  multiplied by $-\alpha$ and $-\beta$, respectively. Therefore, the smooth
  case corresponds to $\beta=-1$, for which the tangential velocity is
  unchanged in the collision. If, in addition, $\alpha=1$ the collision is
  elastic. The completely inelastic case corresponds to $\alpha=0$ ($\beta=0$)
  for the normal (tangential) component}.
\bibitem[{not({\natexlab{b}})}]{note7}
\bibinfo{note}{There are some formal analogies between $\Phi(T^*,\theta)$ and
  $\Psi(T^*,\theta)$, which may be surprising at first sight. From the kinetic
  equation, the evolution equations for the translational and rotational
  temperatures, $T_{t}$ and $T_{r}$, are derived and they are quite different
  (see Ref.~\cite{Vega15} for their explicit expressions). However, both
  quantities contribute to the total temperature $T$, which is their average,
  and to their ratio $\theta$. The formal similarities between $\Phi$ and
  $\Psi$ stem from this ``mixing''}.
\bibitem[{not({\natexlab{c}})}]{note12}
\bibinfo{note}{Since we are interested in cooling experiments, we focus on the
  region of the $(T^{*},\theta)$ plane in which $\Phi<0$}.
\bibitem[{not({\natexlab{e}})}]{note8}
\bibinfo{note}{Note that $\Phi_{2}$ is negative definite, so that its absolute
  value decreases with increasing $\theta$}.
\bibitem[{not({\natexlab{f}})}]{note13}
\bibinfo{note}{Note that, since both $\Phi_2$ and $\Phi_3$ are proportional to
  $\sqrt{T^*}$, $\widetilde{\theta}$ in Eq.~\eqref{thetamin} is independent of
  $T^*$}.
\bibitem[{not({\natexlab{g}})}]{note9}
\bibinfo{note}{Note that $\widetilde{\theta}$ is bounded from above by its
  maximum possible value $\widetilde{\theta}_{\max}=2-3\kappa$, to which it
  tends as $\alpha\to 1$ for all $\beta$. In the particular case
  $\kappa=\frac{2}{5}$, we have $\widetilde{\theta}_{\max}=\frac{4}{5}$}.
\bibitem[{not({\natexlab{h}})}]{note10}
\bibinfo{note}{For longer times, eventual crossings are unobservable since the
  gas has almost reached the steady state. The curves in Fig.\ \ref{phase_diag}
  have been obtained by numerically integrating the nonlinear set of equations
  (\ref{A1}) over a sufficiently fine grid of initial conditions}.
\bibitem[{not({\natexlab{i}})}]{note11}
\bibinfo{note}{With the only exception of the region $\beta\approx 0$, in which
  we know that the angular kurtosis and the translational-rotational
  correlations are more important~\cite{Vega15}}.
\end{thebibliography}

\end{document}